# Experimental demonstration of hybrid improper ferroelectricity and presence of abundant charged walls in (Ca,Sr)$_3$Ti$_2$O$_7$ crystals

Yoon Seok Oh[1,2,†], Xuan Luo[3], Fei-Ting Huang[1,2], Yazhong Wang[1,2], and Sang-Wook Cheong[1,2,3*]

[1]*Rutgers Center for Emergent Materials, Rutgers University, Piscataway, NJ 08854, USA*
[2]*Department of Physics & Astronomy, Rutgers University, Piscataway, NJ 08854, USA*
[3]*Laboratory for Pohang Emergent Materials, Pohang University of Science and Technology, Pohang 790-784, Korea*
[†]*Present address: Department of Physics, Ulsan National Institute of Science and Technology (UNIST), Ulsan 689-798, Korea*
*email: sangc@physics.rutgers.edu.

**Standing on successful first principles predictions for new functional ferroelectric materials, a number of new ferroelectrics have been experimentally discovered. Utilizing trilinear coupling of two types of octahedron rotations, hybrid improper ferroelectricity has been theoretically predicted in ordered perovskites and the Ruddlesden-Popper compounds (Ca$_3$Ti$_2$O$_7$, Ca$_3$Mn$_2$O$_7$, and (Ca/Sr/Ba)$_3$(Sn/Zr/Ge)$_2$O$_7$). However, the ferroelectricity of these compounds has never been experimentally confirmed and even their polar nature has been under debate. Here we provide the first experimental demonstration of room-temperature switchable polarization in the bulk crystals of Ca$_3$Ti$_2$O$_7$ as well as Sr-doped Ca$_3$Ti$_2$O$_7$. In addition, (Ca,Sr)$_3$Ti$_2$O$_7$ is found to exhibit an intriguing ferroelectric domain structure resulting from orthorhombic twins and (switchable) planar polarization. The planar domain structure accompanies abundant charged domain walls with conducting head-to-head and insulating tail-to-tail configurations, which exhibit two-order-of-magnitude conduction difference. These discoveries provide new research opportunities not only on new stable ferroelectrics of Ruddlesden-Popper compounds, but also on meandering conducting domain walls formed by planar polarization.**

There have been numerous attempts of computational materials design based on first principles calculations for new functional materials[1, 2]. A large number of ferroelectric/piezoelectric materials have been computationally predicted[3-7], and some of them were experimentally confirmed[5, 8-13]. For example, the presence of ferroelectricity and strong coupling between magnetism and ferroelectricity were theoretically predicted in $EuTiO_3$[3] and $FeTiO_3$[4]. The polar transition of the compounds was experimentally confirmed[8, 9]. Some of half-Heusler semiconductors are predicted to be new piezoelectrics with large polarization[7]. The theoretical prediction of stabilizing ferroelectricity in strained $Sr_{n+1}Ti_nO_{3n+1}$ ($n \geq 3$)[6] was also experimentally confirmed in biaxially-strained films, which exhibit switchable polarization at low temperatures[10] and the $Sr_{n+1}Ti_nO_{3n+1}$ films with large $n$ such as $SrTiO_3$, corresponding to the $n=\infty$ member, do show ferroelectricity at room temperature[11-13].

Geometric ferroelectrics are improper ferroelectrics where geometric structural constraints, rather than typical cation-anion paring, induce ferroelectric polarization[14]. Hybrid improper ferroelectricity, one kind of geometric ferroelectricity, results from the combination of two or more of non-ferroelectric structural order parameters, and was predicted for a number of compounds including double-layered orthorhombic $Ca_3Ti_2O_7$, $Ca_3Mn_2O_7$, and $(Ca/Sr/Ba)_3(Sn/Zr/Ge)_2O_7$[15-17]. In fact, a few orthorhombic $A_3B_2O_7$ compounds were already known to form in a polar structure, but switching of polarization was never reported for these compounds. Fennie and his company, using first principles calculations, predicted that $Ca_3Ti_2O_7$ and $Ca_3Mn_2O_7$ have too high energy barrier to switch polarization but $(Ca/Sr/Ba)_3(Sn/Zr/Ge)_2O_7$ will have low-enough energy barrier for switchable polarization[17]. However, comparing with other ferroelectric energy barrier (e.g. 20 meV of $BaTiO_3$[18], 30 meV of $PbTiO_3$[19], 25 meV of hexagonal $RMnO_3$ ($R$ = rare earths)[20, 21]), the calculated switching barriers 200 meV of $Ca_3Mn_2O_7$ and ~100 meV of $(Ca/Ba)_3(Sn/Zr)_2O_7$ are still too large to switch polarization[15, 17]. Primarily because of this reason, the validity of hybrid improper ferroelectricity in $A_3B_2O_7$ has been highly under debate[22-24].

Here we report the first experimental demonstration of hybrid improper ferroelectricity in the bulk single crystals of $(Ca,Sr)_3Ti_2O_7$. Electric polarization vs. electric field $P(E)$ hysteresis loops clearly show the existence of switchable polarization with a unexpectedly-low switching electric field. Moreover, in-plane piezo-response force microscope (IP-PFM) images reveal intriguing ferroelectric domain structures comprising abundant meandering charged domain walls. In order to understand the origin of the unexpectedly-low switching electric field and the abundant charged walls, we propose classification of eight types of ferroelectric and four types of ferroelastic domain walls using crystallographic symmetry. Based on the classification, we suggest that individual switching of elementary tilting modes results in the low switching electric field. In addition, the charged domain walls with conducting head-to-head and insulating tail-to-tail configurations exhibit two-order-of-magnitude conduction difference as well as angle-dependent conductivity. These discoveries provide new research opportunities not only on hybrid improper ferroelectricity in Ruddlesden-Popper compounds, but also on abundant conducting domain walls stemming from planar polarization.

$Ca_3Ti_2O_7$ (and low-Sr-doped $Ca_3Ti_2O_7$) forms in an orthorhombic structure with $A2_1am$ space group as shown in Figure 1a and 1b. The layered perovskite structure consists of perovskite (P) block and Rock-salt (R) block. The P block is built up of two layers of corner sharing $TiO_6$ octahedra, and Ca/Sr ions locate at the body center of the P block. In-between the adjacent P blocks, O and Ca/Sr ions are arranged in the R block. As shown in Fig. 1a, the $TiO_6$ octahedra rotate alternatively in the (001) plane around [001], which is represented by $a^0a^0c^+$ in the Glazer notation (see Supplementary Figure S1c)[25]. In addition, the $TiO_6$ octahedral tilting around [100] axis, $a^-a^-c^0$ tilting in the Glazer notation, results in off-centered apical oxygen ions (dark green) (Figure 1a and Supplementary Figure S1e). The simultaneous presence of $TiO_6$ tilting and rotation leads to Ca/Sr displacements along [100] (red and blue arrows in Fig. 1a and b) and also non-zero electric polarization along [100], according to the scenario of hybrid improper ferroelectricity[15-17].

We were able to grow high-quality single crystals of $(Ca,Sr)_3Ti_2O_7$, and confirmed the presence of switchable ferroelectric polarization in the orthorhombic crystals at room temperature. The orthorhombic structural distortion in $(Ca,Sr)_3Ti_2O_7$ results in orthorhombic twins. Figure 1c and d show photographic and circular differential interference contrast (cDIC) images of the cleaved (001) surface of a $Ca_{2.46}Sr_{0.54}Ti_2O_7$ single crystal, respectively. Dark-bright contrast in the cDIC image (Fig. 1d) indicates the presence of orthorhombic twins. Under linearly polarized light, the dark-bright contrast of these twins is reversed by changing the relative angle between polarizer and analyzer (Supplementary Figure S2). Twin boundaries are along [110], and polarization is supposedly along [100]. Thus, in order to switch polarization in both types of twin domains at the same time, an electric field is applied along [110], *i.e.* parallel to the twin boundaries. Figure 1e shows electric polarization vs. electric field $P(E)$ hysteresis loops of $Ca_{3-x}Sr_xTi_2O_7$ ($x$=0, 0.54, and 0.85) single crystals (see Supplementary Figure S3 for more details), demonstrating not only the presence but also the switching of remanent polarization at room temperature. The net remanent electric polarizations ($P_n = \sqrt{2} \cdot P_{r,exp}$) along the [100] direction of $x$=0, 0.54, and 0.85 are estimated to be 8, 4.2, and 2.4 $\mu C/cm^2$, respectively. This decreasing net polarization with substitution of larger Sr cation is in contrast with a theoretical prediction in $(Ca/Sr/Ba)_3(Sn/Zr/Ge)_2O_7$[17]. Although the theory didn't predict the doping dependence of polarization in the Sr doped $Ca_3Ti_2O_7$, the mechanisms of hybrid improper ferroelectricity in $Ca_3Ti_2O_7$ and $(Ca/Sr/Ba)_3(Sn/Zr/Ge)_2O_7$ are expected to be identical. Even in the Sr doped $Ca_3Ti_2O_7$, a larger-cation substitution of small cations at the P block likely increases the net polarization[17]. Distinctly different coordination numbers of $A$-cation sites at the P and R blocks and the large difference of Ca/Sr ion size should lead to the preferential substitution of Sr ions in the P block[26, 27]. Enhancement of orthorhombicity as a function of Sr concentration (Supplementary Figure S4c) strongly supports that larger Sr ions preferentially occupy at the P block and decrease both $a^0a^0c^+$ rotation and $a^-a^-c^0$ tilting. Thus, this decrease of $a^0a^0c^+$ rotation and $a^-a^-c^0$ tilting compensates the effect of Ca/Sr ionic ordering, and results in the decrease of the net polarization in $Ca_{3-x}Sr_xTi_2O_7$ with increasing Sr doping.

In-plane piezo-response force microscope (IP-PFM) images reveal intriguing planar polar domains on the cleaved (001) surfaces of $Ca_3Ti_2O_7$ and $Ca_{2.46}Sr_{0.54}Ti_2O_7$ (Fig. 2a and b, respectively). The gray arrows in the bottom panels of Fig. 2a and b depict ferroelastic (FA) domain walls (DWs), which are identified as orthorhombic twin boundaries in the cDIC images. Our IP-PFM experiment confirms that polarizations orient with a ~45° angle to the FA DWs, thus the FA DWs form 90° ferroelectric (FE) DWs. In the crystallography of ferroelastics, twin domains should be mirror-symmetric with respect to a twin boundary[28] ($\theta_{FA}$=+45° and -135° in Fig. 2c). However, the mirror-symmetric twin domains in a ferroelectric form charged (head-to-head or tail-to-tail) walls with large electrostatic energy. To reduce the electrostatic energy cost, uncharged (head-to-tail) twin boundaries ($\theta_{FA}$=-45° and +135° in Fig. 2c) are expected to be stabilized, similar with what happens in twin boundaries of ferroelectric $BaTiO_3$[29] and PZT[30]. Based on this similarity[28-30], we propose four possible configurations of the FA DWs: head-to-head, tail-to-tail, head-to-tail with the counter-clock-wise rotation, and head-to-tail with the clock-wise rotation in Fig. 2c. In this picture, the formation of the FA DWs involves switching of individual $a^-a^0c^0$ and

$a^0a^-c^0$ tilting's (refer Fig. 2e). Here, we suggest a model to explain the stabilized 90° FE DWs (FA DWs). Since $a^-a^-c^0$ is a collective mode of $a^-a^0c^0$ and $a^0a^-c^0$, the order parameter of $a^-a^-c^0$, $Q_{a^-a^-c^0}$, can be described as $Q_{a^-a^-c^0} = Q_{a^-a^0c^0} + Q_{a^0a^-c^0}$ [31]. Then, the relevant free energy can be expressed as $F = \alpha Q_{a^-a^-c^0}Q_{a^0a^0c^+}P_{[110]_{pc}} = \alpha_1 Q_{a^-a^0c^0}Q_{a^0a^0c^+}P_{[100]_{pc}} + \alpha_2 Q_{a^0a^-c^0}Q_{a^0a^0c^+}P_{[010]_{pc}}$, where α, α1, and α2 are coupling constant and $P_{[100]} = P_{[110]_{pc}} = P_{[100]_{pc}} + P_{[010]_{pc}}$ [31]. This means that, as long as the $a^0a^0c^+$ mode is unchanged, switching of individual $a^-a^0c^0$ and $a^0a^-c^0$ results in switching of $P_{[100]_{pc}}$ and $P_{[010]_{pc}}$, respectively, and stabilizes the 90° FE DWs. Referring Fig. 2e, all of FA DWs (90° FE DWs) are accompanied by reversal of either $a^-a^0c^0$ or $a^0a^-c^0$ tilting, but reversing neither $a^-a^-c^0$ nor $a^0a^0c^+$ cannot induce a FA DW.

    Within one FA domain, a number of polar domains exist with meandering 180° FE DWs. The existence of the 180° FE DWs is inconsistent with the results in ref. 32, where 180° FE DW is expected unstable. In Addition, our IP-PFM experiment after complete poling (see Supplementary Figure S5) reveals that switching of ferroelectric polarization in (Ca,Sr)$_3$Ti$_2$O$_7$ is accompanied by 180° switching of the polarization. In order to explain the formation of the 180° FE DWs and the 180° switching of the polarization, we introduce two hypotheses for the 180° FE DWs. First, adjacent domains should be inversion-symmetric with respect to a 180° FE DW. Second, the DW of two inversion-symmetric domains should form without any distortion of TiO$_6$ octahedral cages. Then, local structure of a 180° FE DW, shown in Fig. 2d, is identified by imposing an inversion center at the center of the orthorhombic unit cell. We can expect eight different DW angles ($\theta_{FE}$=+135°, +90°, +45°, 0°, -45°, -90°, -135°, and ±180°) with respect to a polarization direction can be the basis to form the meandering 180° FE DWs (Fig. 2d). $\theta_{FE}$ denotes the angle between a polarization and a 180° FE DW under a fixed coordinate (see Fig. 2d). It's supposed that the adjacent domains with respect to a 180° FE DW are associated with the opposite $a^-a^-c^0$ tilting's but an identical $a^0a^0c^+$ rotation. Thus, switching of $a^0a^0c^+$ rotational mode doesn't contribute to forming 180° FE DWs and inducing 180° switching. However, according to ref. 15, polarization switching is involved with switching of the $a^0a^0c^+$ mode. Switching the $a^0a^0c^+$ rotation mode, which is associated with a very deep energy well, was predicted to lead to an inaccessibly-large switching coercive field[15], but our measured coercive field value of 120kV/cm in (Ca,Sr)$_3$Ti$_2$O$_7$ is comparable with that of other proper ferroelectrics; e.g. 200 kV/cm of BaTiO$_3$

films[33] and 140 kV/cm of BiFeO$_3$ films[34]. Thus, it's possible that the $a^0a^0c^+$ rotation mode may not be necessarily required to switch polarization in hybrid improper ferroelectric. Instead, individual $a^-a^0c^0$ and $a^0a^-c^0$ tilting's can contribute to switch $P_{[100]_{pc}}$ and $P_{[010]_{pc}}$ components of the net polarization $P_{[100]} = P_{[110]_{pc}}$, respectively, similar with what happens in BiFeO$_3$/LaFeO$_3$ superlattices[31]. This may be the reason why the polarization of (Ca,Sr)$_3$Ti$_2$O$_7$ is switched by a reasonably small and accessible electric field, in contrast with theoretical predictions[15, 17]

Statistical analysis of the angular distribution of the length of 180° FE DWs from the IP-PFM images of Ca$_3$Ti$_2$O$_7$ and Ca$_{2.46}$Sr$_{0.54}$Ti$_2$O$_7$ provides insights in understanding the formation of 180° FE DWs at different angles (Fig. 3a and b). For Ca$_3$Ti$_2$O$_7$ with large domains (Fig. 3a), a good fraction of the 180° FE DWs appear to be confined at $\theta_{FE}=\pm135°$ and $\pm45°$, but 180° FE DWs with $\theta_{FE}=0°$, $\pm90°$, and $\pm180°$ as well as $\theta_{FE}=\pm135°$ and $\pm45°$ are observed in another area of Ca$_3$Ti$_2$O$_7$ (Supplementary Figure S6). Ca$_{2.46}$Sr$_{0.54}$Ti$_2$O$_7$ with small domains (Fig. 3b) exhibits the 180° FE DWs with most of eight angles in the same areal size (60μm×60μm). Therefore, there exists no preferred $\theta_{FE}$ of FE DWs even though different amounts of charge (*i.e.* different electrostatic energy costs) are expected for different $\theta_{FE}$'s. Note that 180° FE DWs with $\theta_{FE}=0°$ and $\pm180°$ are non-charged, but are not particularly favored. We emphasize that the presence of these charged 180° FE DWs in the (Ca,Sr)$_3$Ti$_2$O$_7$ is not consistent with a theoretical predictions[32].

Figure 3c and d illustrate schematics of polarization domains obtained from both horizontal and vertical IP-PFM images of Ca$_3$Ti$_2$O$_7$ and Ca$_{2.46}$Sr$_{0.54}$Ti$_2$O$_7$, respectively. We find that one FA DW does contain all four configurations: head-to-head ($\theta_{FA}=+45°$), tail-to-tail ($\theta_{FA}=-135°$), head-to-tail with the counter-clock-wise rotation ($\theta_{FA}=+135°$), and head-to-tail with the clock-wise rotation ($\theta_{FA}=-45°$), even though $\theta_{FA}=+45°$ and $-135°$ are likely involved with a large cost of electrostatic energy. Note that the charge densities of FA DWs with $\theta_{FA}=+45°$, $-135°$, and $+135°/-45°$ are supposedly same with those of $\theta_{FE}=+45°/+135°$, and $-135°/-45°$, and $0°/\pm180°$ of 180° FE DWs, respectively. We find that the FA DWs (diamond symbols in Fig. 3a and b) do not exhibit any preferred angle, indicating no preference of non-charged FA DWs with $\theta_{FA}=+135°$ or $-45°$. And, the Sr doping produces more FA DWs as well as 180° FE DWs. In the same areal size (60μm×60μm), the total length ($\approx$655μm) of FA DWs in Ca$_{2.46}$Sr$_{0.54}$Ti$_2$O$_7$ is six times larger than that ($\approx$108μm) in Ca$_3$Ti$_2$O$_7$. As we discussed earlier, the Sr substitution leads to the reduction of

$a^0a^0c^+$ rotation and $a^-a^-c^0$ tilting. As a consequence, the elastic energy cost for FA DWs also decreases, and thus, $Ca_{2.46}Sr_{0.54}Ti_2O_7$, compared with pure $Ca_3Ti_2O_7$, exhibits more FA DWs.

A unique feature of the domain structure of as-grown $Ca_{3-x}Sr_xTi_2O_7$ bulk single crystals is the existence of abundant charged DWs even though they usually accompany a large electrostatic energy cost. Due to this large electrostatic energy cost, in bulk ferroelectric crystals, charged FE DWs have been observed only in very limited cases. In hexagonal $R$MnO$_3$ single crystals, the non-trivial topology of FE domains results in charged FE DWs[35, 36]. In the presence of external electric fields, charged FE DWs have been observed in bulk single crystals such as $BaTiO_3$ as a metastable state[37]. Note that bulk crystals of $BiFeO_3$ rarely exhibit charged FE DWs[38], even though $BiFeO_3$ films[39] do exhibit charged FE DWs, probably due to significant structural disorder. We have studied conducting properties of the charged DWs in a slightly-oxygen-deficient $Ca_{2.44}Sr_{0.56}Ti_2O_{7-\delta}$ single crystal using IP-PFM and conductive atomic force microscopy (cAFM) (Fig. 4a and b). We find that DWs, compared with domains, can exhibit large conduction, and the magnitude of conduction at DWs varies significantly, depending on DW angles. From the conduction map in Fig. 4b, it is rather evident that positively-charged $\theta_{FE}=+90°$ FE DWs exhibit large conduction, and negatively-charged $\theta_{FE}=-90°$ FE DWs show poor conduction. These results are consistent with the n-type semiconducting nature of $(Ca,Sr)_3Ti_2O_{7-\delta}$ in the sense that positively-charged DWs attract n-type charge carriers (so accompany large conduction) and negatively-charged DWs expel n-type charge carriers (so accompany poor conduction). Two domains, shown in the inset of Fig. 4c, are selected to obtain the exact angular dependence of normalized conductance $G^{norm}(\theta_{FE})=(I_{wall}(\theta_{FE})-I_{bulk})/I_{bulk}$ (green closed and opened circles in Figure 4c), where $I_{wall}$ and $I_{bulk}$ are the measured current at DWs and in domains, respectively. The solid diamonds in Fig. 4c indicate $(I_{wall}(\theta_{FA})-I_{bulk})/I_{bulk}$ at FA DWs. The angular dependence of $(I_{wall}-I_{bulk})/I_{bulk}$ for FE DWs is in accordance with that of the calculated DW conductance (black line in Fig. 4c) using a modified equilibrium model proposed by Meier *et al.* to explain the conduction in charged walls in ErMnO$_3$[35]. In our calculation of DW conductance, we used the measured values ($P$=4.2 μC/cm$^2$, $\varepsilon$=30$\varepsilon_0$, and $w$=50 nm) of polarization, dielectric constant and domain wall width, and the carrier density of $n_0$=8.5×10$^{18}$cm$^{-3}$ that result in the best fit. Note that this carrier density is similar with that in oxygen-reduced SrTiO$_3$ at the same annealing temperature range[40]. This good agreement indicates that band reconstruction accompanied by charge accumulation leads to the variable conductance at charged FE and FA DWs established by planar polarization. The conduction at uncharged $\theta_{FA}$=-45° and +135° FA DWs is comparable with that of the neighboring domains as shown in the inset of Fig. 4d. The conductance at the positively-charged $\theta_{FA}$=+45° and negatively-charged $\theta_{FA}$=-135° FA DWs is three times larger and ten times smaller than the domain's conductance, respectively. Note that, compared with $\theta_{FE}$=+45° and +135° FE DWs, $\theta_{FA}$=+45° FA DWs show less conductance even though they have the same localized charge density. This difference may result from the fact that Ti ions sit right at $\theta_{FE}$=+45° and +135° FE DWs, while Ca/Sr ions locate at $\theta_{FA}$=+45° FA DWs, so the band gap at $\theta_{FE}$=+45° and +135° FE DWs is smaller than that at $\theta_{FA}$=+45° FA DWs. Local *I-V* characteristics at FE DWs (Fig. 4d) demonstrate that there exists a Schottky-type barrier between cAFM tip and the specimen, and the $\theta_{FE}$=+45° (head-to-head) FE DW exhibits hundred-times larger conductance than the $\theta_{FE}$=-135° (tail-to-tail) FE DW does with a positive forward bias. Emphasize that this large conduction difference is almost two order of magnitude larger than that in other materials such as hexagonal $R$MnO$_3$[35] and BaTiO$_3$[37].

In summary, we have successfully grown high-quality single crystals of Ca$_{3-x}$Sr$_x$Ti$_2$O$_7$ (*x*=0, 0.54, 0.85) and demonstrated improper ferroelectricity of the system. Suitable chemical stability enables to grow bulk single crystals using a traditional optical floating zone method. Switchable planar polarization with clear *P*(*E*) hysteresis loops has been observed in highly insulating as-grown crystals. In contrast with a theoretical prediction, the measured coercive field 120kV/cm for switching polarization, involving switching of oxygen-cage-rotational modes, is accessibly small. In order to explain the unexpected-low switching electric field for abundant 180° FE DWs and 90° FE DWs (FA DWs), we have introduced a model in which switching of individual a$^-$a$^0$c$^0$ and a$^0$a$^-$c$^0$ tilting's paves the way for the switchable polarization. This unexpected discovery should stimulate further scientific investigations for more explicit experimental demonstrations of the Glazer rotations in the intriguing FE domains and open possible utilization of hybrid improper ferroelectrics for real technological applications. The two dimensionality of the *n*=2 Ruddlesden-Popper structure of (Ca,Sr)$_3$Ti$_2$O$_7$ leads to intriguing planar ferroelectric domains with eight types of ferroelectric and four types of ferroelastic domain walls. Six out of the eight types of ferroelectric domain walls exhibit 180° charged domain wall configurations, and two out of the four types of ferroelastic domain walls show 90° charged wall configurations. Furthermore, in a slightly-reduced $Ca_{2.46}Sr_{0.54}Ti_2O_{7-\delta}$ single crystal, domain wall conductivity modulates with domain wall angle, associated with the angular distribution of charge density at domain walls, and head-to-head and tail-to-tail 180° ferroelectric domain walls exhibit ~100 times conduction difference in cAFM measurements. In terms of conductivity and rectification, it's expected to show better or at least comparable functionality than the reported ferroelectric compounds do[37, 39, 41, 42]. Therefore, our discovery of abundant charged walls in $Ca_{3-x}Sr_xTi_2O_7$ single crystals, some of which are conducting and some others of which are insulating, provides new insights into understanding and engineering conducting domain walls of planar ferroelectrics.

## Methods

**Sample Preparations.** $Ca_{3-x}Sr_xTi_2O_7$ single crystals were grown using an optical floating zone method. For polycrystalline $Ca_{3-x}Sr_xTi_2O_7$ feed rods, stoichiometric CaO, SrO, and $TiO_2$ were mixed, ground, pelletized, and sintered at 1200°C for 15 hours. For oxygen deficient single crystals, as-grown single crystals were annealed at 800°C in $N_2$ flow.

**Measurements.** Electric polarization vs. electric field, $P(E)$, hysteresis loops were measured at room temperature in silicon oil. The electric fields at the frequency of 260Hz for $E$//[110] and 1kHz for $E$//[001] were applied using a programmable function generator (DS340) and a high voltage amplifier, and current vs. $E$ data were recorded using an oscilloscope (TDS1010). Polarization was obtained from integrating the current vs. $E$ data. In order to separate leakage and dielectric contributions, the so-called positive-up-negative-down (PUND) method was employed[43]. X-ray diffraction (XRD) experiments were performed using a Rigaku D/Max-RB x-ray diffractometer with Cu $K_\alpha$ radiation.

The IP-PFM experiments were performed on the cleaved (001) surfaces of $Ca_3Ti_2O_7$ and $Ca_{2.46}Sr_{0.54}Ti_2O_7$. A large Au electrode was deposited on a (001) surface as the bottom electrode. Small AC voltage was applied to the bottom electrode and a conducting AFM tip was grounded as shown in Fig. 1f. Thus, the AC voltage was applied perpendicular to the polarization direction. The IP-PFM signal in Figure 2a, 2b, and 4a originates from the shear piezoelectric response $d_{25}$.

Polar $(Ca,Sr)_3Ti_2O_7$ structure has 2mm symmetry, where the 2-fold axis is parallel to the [100] axis. The symmetry of polar $(Ca,Sr)_3Ti_2O_7$ allows the following piezoelectric tensor.

$$d_{ij} = \begin{pmatrix} d_{11} & d_{12} & d_{13} & 0 & 0 & 0 \\ 0 & 0 & 0 & d_{24} & 0 & 0 \\ 0 & 0 & 0 & 0 & d_{25} & 0 \end{pmatrix}$$

where index $i, j = 1, 2, 3$ correspond to [100], [010], [001] axes. The shear piezoelectric response $d_{25}$ of $Ca_3Ti_2O_7$ and low-Sr-doped $Ca_3Ti_2O_7$ with the 2mm symmetry is non-zero. Using the lateral force mode of a SPM controller (Digital Instruments Nanoscope IIIa) and a lock-in amplifier (SR830), we extracted the in-plane piezoelectric response perpendicular to the AFM cantilever's long axis.

The cAFM was measured on the cleaved (001) surfaces of the oxygen deficient $Ca_{2.46}Sr_{0.54}Ti_2O_7$ single crystal with a current amplifier (SR570).

**Statistical Analysis.** After drawing domains wall lines over the experimental IP-PFM image by hands, the lines were digitized using OriginLab OriginPro 9.1. Distance and angle between neighboring data points in the lines were calculated with respect to fixed coordinates and an origin. The coordinates for twin domains were rotated by 90° to each other to compensate the polarization 90° rotation effect. For the statistical analysis of the angular distribution of domain wall length, domain wall lines were segmented for each 6° interval in $-180° \leq \theta < 180°$.

**Acknowledgements**

The work at Rutgers University was supported by the NSF under Grant No. NSF-DMR-1104484, and that at Postech by the Max Planck POSTECH/KOREA Research Initiative Program [Grant No. 2011-0031558] through NRF of Korea funded by MEST.

**Author contributions**

Y.S.O. carried out $P(E)$, IP-PFM, IP-PFM before/after poling, cAFM, $I$-$V$ measurements, and statistical analysis and conceived the hypothetical model. X.L. synthesized single crystals and performed XRD. F.T.H. performed the structure refinement. Y.S.O. and S.W.C. analyzed the data and wrote the manuscript. S.W.C. initiated and supervised the research.

**Competing financial interests**

The authors declare no competing financial interests.

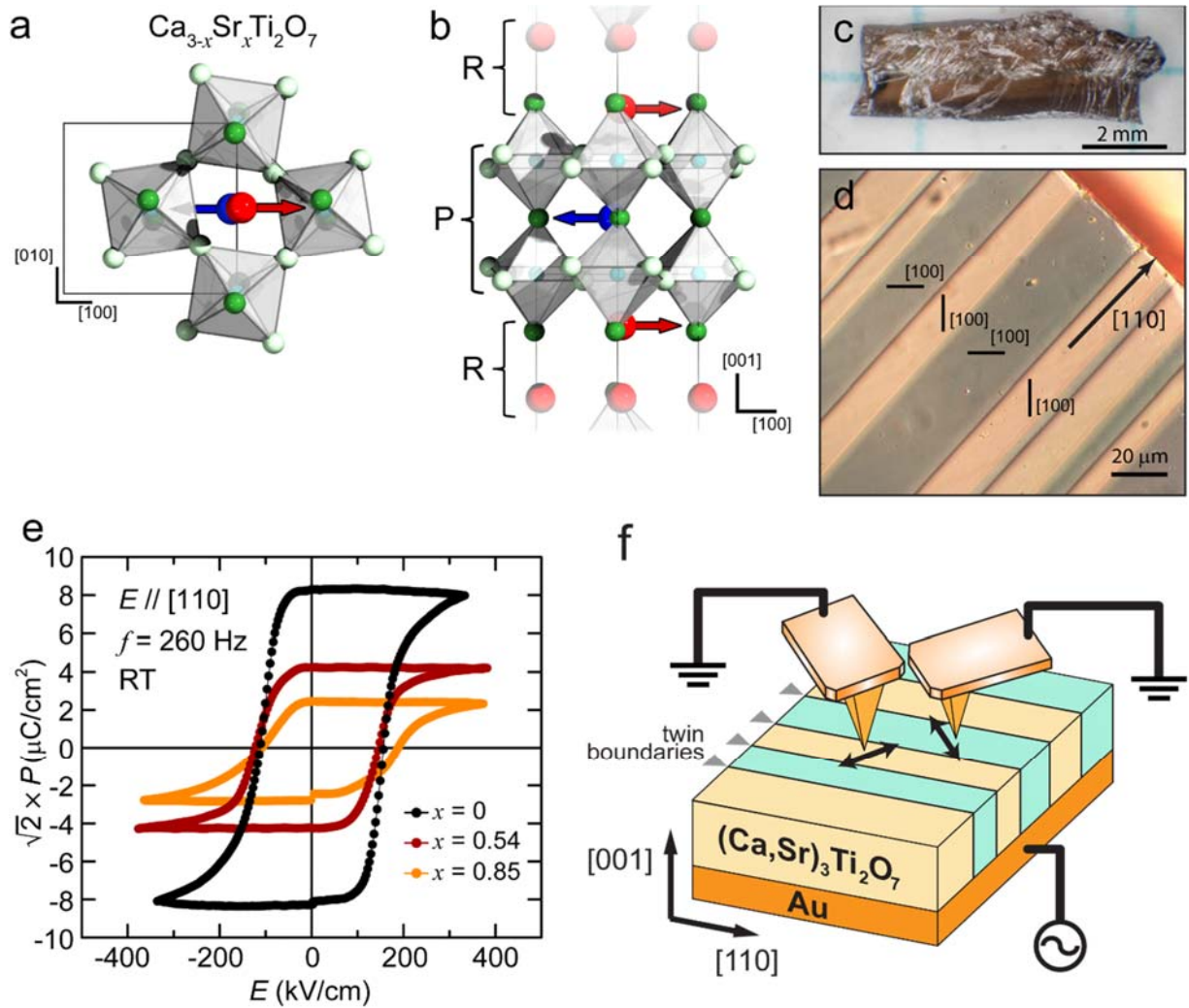

**Figure 1. Planar electric polarization of $Ca_{3-x}Sr_xTi_2O_7$ single crystals at room temperature. a-b**, Crystallographic structure of $Ca_{3-x}Sr_xTi_2O_7$ with orthorhombic $A2_1am$ space group. The red (blue), light green (dark green), and light-blue spheres represent Ca/Sr in Rocksalt R block (perovskite P block), planar (apical) oxygen O of $TiO_6$ octahedra, and Ti at center of the $TiO_6$ octahedra, respectively. The red and blue arrows depict Ca/Sr displacement along [100]//$a$ in the R and the P block, respectively. **c,** Photographic image of a cleaved (001) surface of a $Ca_{2.46}Sr_{0.54}Ti_2O_7$ single crystal. **d,** Circular differential interference contrast (cDIC) image of the cleaved (001) plane of a $Ca_{2.46}Sr_{0.54}Ti_2O_7$ single crystal at room temperature. This cDIC image

shows the presence of orthorhombic twins. Black lines display the orientation of the *a*-axis (//[100]) in orthorhombic twins. **e,** Electric polarization vs. electric field, *P*(*E*), hysteresis loops of $Ca_{3-x}Sr_xTi_2O_7$ (*x*=0, 0.54, 0.85) single crystals at room temperature with electric fields applied along the [110] direction (*i.e.* twin boundaries) at frequency *f*=260Hz. **f,** Schematic picture of our IP-PFM measurement. Cyan and yellow stripes distinguish twin domains like in **d**. Gray arrows depict the twin boundaries.

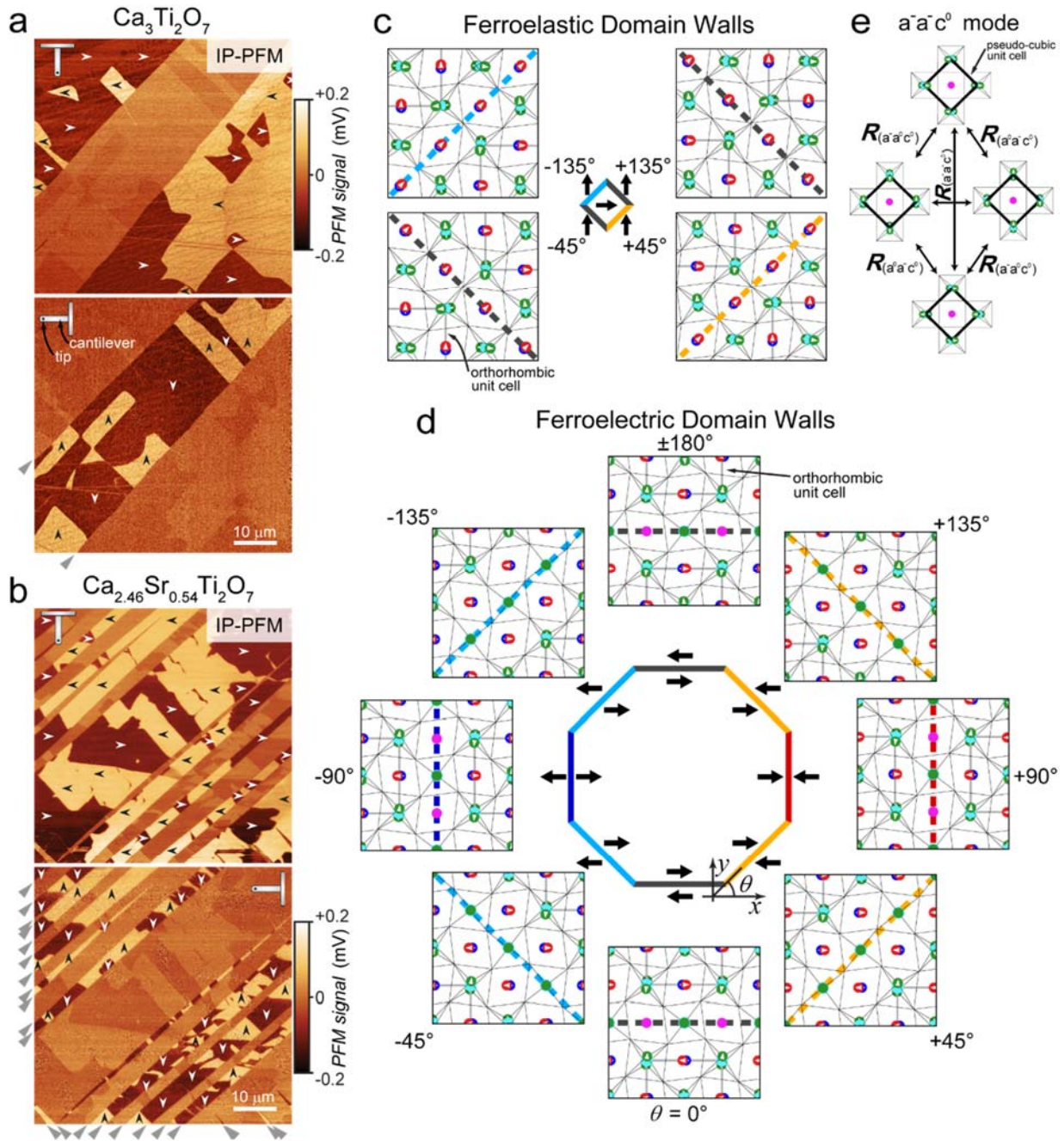

**Figure 2. Abundant ferroelectric (FE) and ferroelastic (FA) domain walls (DWs). a-b**, In-plane piezo-response force microscope (IP-PFM) images of the cleaved (001) surfaces of $Ca_3Ti_2O_7$ and $Ca_{2.46}Sr_{0.54}Ti_2O_7$ crystals at room temperature. FA DWs (orthorhombic twin boundaries) are oriented along the diagonal direction of the XY scanning axes, and the AFM cantilever's long axis

is oriented along the vertical (horizontal) direction as depicted in the top (bottom) panels of **a** and **b**. White and black arrows show the directions of planar polarization, and gray arrows on the outside of the panels point the FA DWs. **c-d,** The schematic pictures in the middle depict possible DW angles with respect to the polarization direction (black arrows). Edges of diamond (octagon) in the middle schematics and thick dashed lines in the crystallographic cartoons correspond to FA (FE) DWs. Different colors of DWs represent different amounts of localized charges due to polarization surface charges. DW angle ($\theta$) is defined by the angle of the DW with respect to the coordinates shown in **d**. The positive and negative signs of $\theta$ in **d** indicate head-to-head and tail-to-tail configurations, respectively. For FA DWs in **c**, $\theta_{FA}$=+135° and -45° correspond to the counter-clock-wise and clock-wise rotation of head-to-tail polarization at the DWs, respectively. Panels with crystallographic cartoons show local distortions for four different DW angles of FA DWs and eight different DW angles of FE DWs. Red/blue, pink, cyan, green solid circles indicate displaced Ca/Sr, non-displaced Ca/Sr, Ti, and O ions, respectively (consistent with those in Figure 1a and b), and the planar O ions are omitted. The arrows in the circles and gray solid lines in the crystallographic cartoons illustrate the directions of atomic distortions and orthorhombic unit cell boundaries, respectively. **e,** Schematics of switching pathways between various oxygen-cage rotations of the $a^-a^-c^0$ modes. $\boldsymbol{R}_{(a\text{-}a0c0)}$, $\boldsymbol{R}_{(a0a\text{-}c0)}$, and $\boldsymbol{R}_{(a\text{-}a\text{-}c0)}$ are switching operators of $a^-a^0c^0$, $a^0a^-c^0$, and $a^-a^-c^0$ modes. The thick black lines are pseudo-cubic unit cell boundaries.

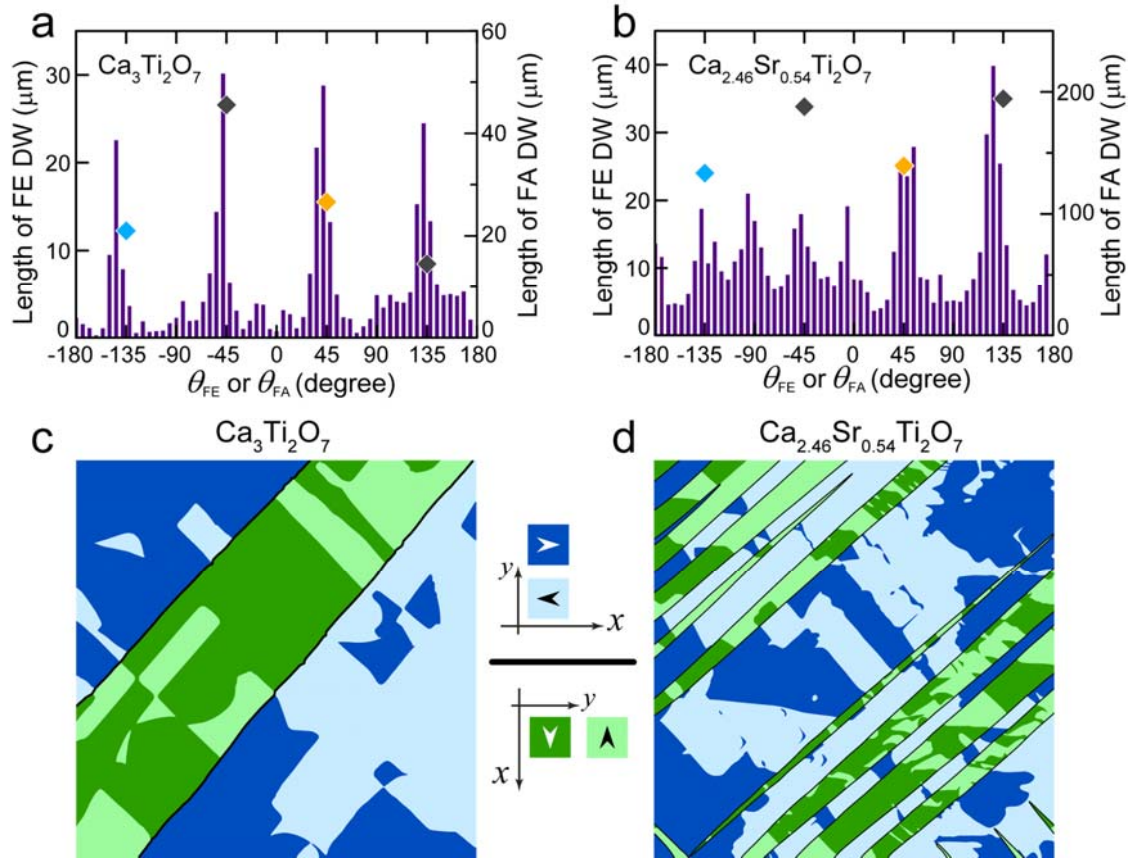

**Figure 3. Statistical analysis for the angular distribution of ferroelectric (FE) and ferroelastic (FA) domain walls (DWs). a-b,** Statistical distribution of the length of FE (purple columns) and FA (solid diamonds) DWs vs. domain wall angle. Purple columns represent the total length of all FE DWs for each $\Delta\theta_{FE}=6°$ interval. Orange, light blue, and gray diamonds depict the total length of FA DWs with $\theta_{FA}=+45°$, -135°, and +135°/-45°, respectively. **c-d,** Illustrations of polar domains in 60μm×60μm areas, obtained from experimental IP-PFM images on $Ca_3Ti_2O_7$ (Fig. 2a) and $Ca_{2.46}Sr_{0.54}Ti_2O_7$ (Fig. 2b) crystals. Dark (light) blue and dark (light) green colors indicate rightward (leftward) and downward (upward) polar domains, respectively. The coordinates for different twin domains are shown in-between **c** and **d**. Black lines show FA DWs (*i.e.* orthorhombic twin boundaries).

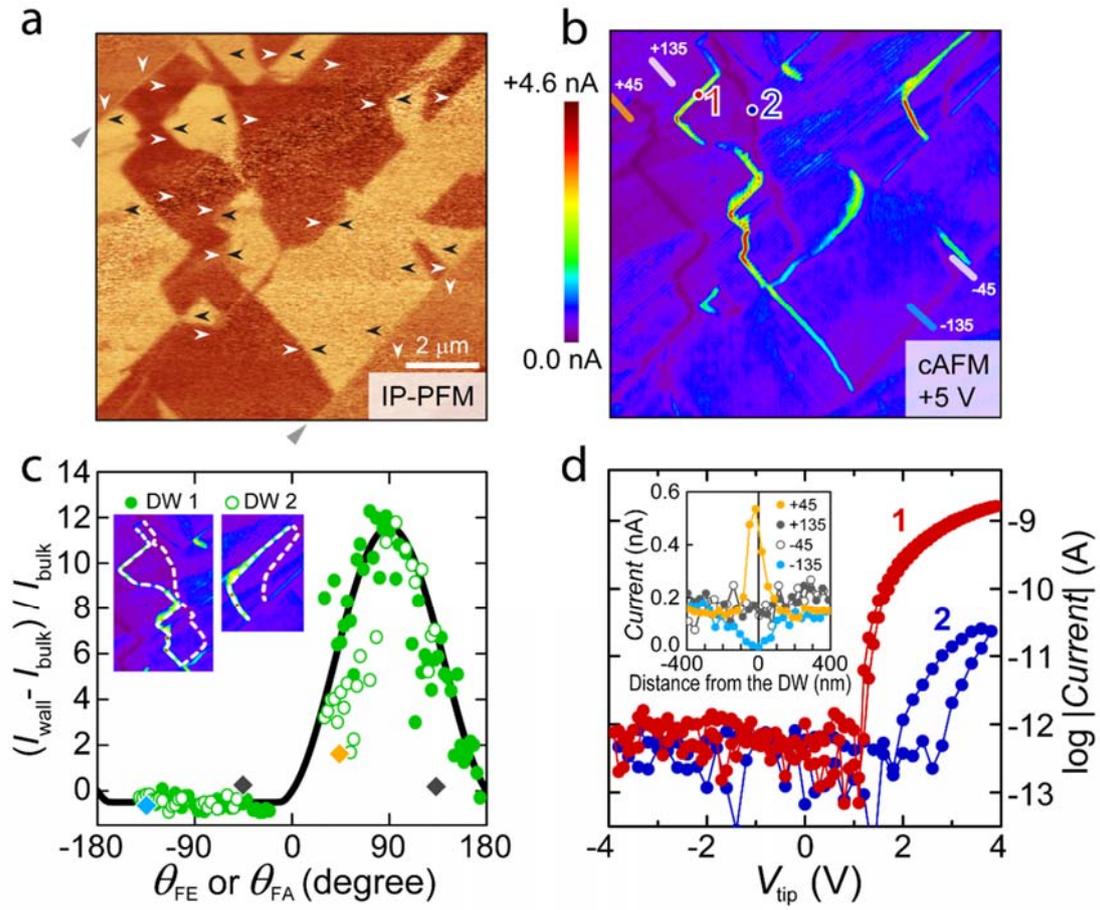

**Figure 4. Angular dependence of the conductance of ferroelectric (FE) and ferroelastic (FA) domain walls (DWs). a,** IP-PFM image of the (001) surface of an oxygen deficient $Ca_{2.46}Sr_{0.54}Ti_2O_{7-\delta}$ crystal at room temperature. The AFM cantilever's long axis is oriented along the vertical direction of the image. **b**, Conductive atomic force microscope (cAFM) image with the voltage of $V_{tip}$=+5V applied to the tip. Color scale depicts the magnitude of the measured cAFM current. **c,** Angular dependence of the conductance of FE (green open/close circles) and FA (colored solid diamonds) DWs. The conductance is normalized as $G^{norm}(\theta)=(I_{wall}(\theta)-I_{bulk})/I_{bulk}$, where $I_{wall}(\theta)$ is the measured current at DW and the averaged current at domain, $I_{bulk}$, is 0.2nA. The selected DWs for the analysis of the angular dependence of the DW conductance are shown with white dashed lines in the inset. Orange, light blue, and gray diamonds depict the normalized

conductance of FA DWs with $\theta_{FA}$=+45°, -135°, and +135°/-45°, respectively. The black curve is the calculated angular dependence of the normalized conductance (see the text for details). **d,** *I-V* curve (red) at one spot on a head-to-head FE DW (the red dot with the number 1 in **b**) and *I-V* curve (blue) at one spot on a tail-to-tail FE DW (the blue dot with the number 2 in **b**). The inset depicts the current profile while a cAFM tip scans across four types of FA DWs. Scanning lines are shown as solid lines in **b**, and the color of data points matches with that of the corresponding scanning line.